 \documentstyle[prl,aps,epsf]{revtex}  %% Journal Style.  
\begin{document}  
\draft  
\preprint{\today}  
  
%% Include following two lines for Journal Style  
  
 \twocolumn[\hsize\textwidth\columnwidth\hsize  %**Journal  
 \csname @twocolumnfalse\endcsname              %**Journal  
  
\title{Quadrupole and octupole softness in the $N=Z$ nucleus $^{64}$Ge}  
\author{ Kazunari Kaneko,$^{1}$ Munetake Hasegawa,$^{2}$ and Takahiro Mizusaki$^{3}$}  
\address{  
$^{1}$Department of Physics, Kyushu Sangyo University,
Matsukadai, Fukuoka 813-8503, Japan\\
$^{2}$Laboratory of Physics, Fukuoka Dental College, Fukuoka 
814-0193, Japan\\
$^{3}$Institute of Natural Sciences, Senshu University, Higashimita, Tama, Kawasaki, Kanagawa, 
214-8580, Japan\\
}  
  
\date{\today}  
\maketitle  
  
\begin{abstract}  
Quadrupole and octupole softness in the even-even $N=Z$ nucleus $^{64}$Ge is studied 
on the spherical shell model basis. We carry out the shell model calculation
 using the pairing plus quadrupole ($QQ$) plus octupole ($OO$) interaction 
 with monopole corrections. 
It is shown that $^{64}$Ge is an unstable nucleus with respect to both the quadrupole
and octupole deformations, which is consistent with the previous discussions
predicting the $\gamma$ softness and octupole instability. 
It is demonstrated that proton-neutron part $Q_{p}Q_{n}$ of the $QQ$ interaction is 
important for the $\gamma$ softness or triaxiality. 

\end{abstract}  
  
\pacs{PACS: 21.60.Cs, 21.60.Ev, 23.20.-g, 27.50.+e}  
  
%%  Include the following line for Journal style  
 ]  %**Journal  
  
\narrowtext  
%\newpage  
\bigskip  

Heavy $N=Z$ nuclei with $A= 56\sim 80$ show strong shape variations such as 
prolate shape, oblate shape, prolate-oblate shape coexistence, and $\gamma$ softness, 
depending on the mass number. 
These nuclei lie in transitional regions from spherical shape 
(e.g., $^{56}$Ni \cite{Rudolph1}) 
to strong prolate deformation (e.g., $^{80}$Zr \cite{Lister}). 
The $N=Z=32$ nucleus $^{64}_{32}$Ge$_{32}$ is known to be
a typical example showing $\gamma$-soft structure in
 $N=Z$ proton-rich unstable nuclei, according to theoretical calculations
 based on the mean-field approximation \cite{Ennis}. 
The calculations predict probable $\gamma$ instability in the ground state,
 and triaxiality in the excited states, i.e.,
  the quadrupole deformation $\beta_{2}\sim 0.22$ and $\gamma\sim 27^{\circ}$. 

Deformed shell model calculations predict that the nucleon numbers 34, 56, 88,
 and 134 are strongly octupole driving in nuclei where the Fermi surface
 lies near single-particle levels with $\Delta l=\Delta j=3$ \cite{Butler}. 
We can expect that nuclei with $N=Z$ near the octupole magic numbers exhibit
 an especially strong octupole effect,
 because neutrons and protons contribute cooperatively. 
This does not necessarily mean a permanent octupole deformation. 
The level pattern of negative-parity states in $^{64}$Ge is not a rotational 
one \cite{Ennis}, and 
the sequence of the $3^{-}$, $5^{-}$, and $7^{-}$ levels is irregularly spaced. 
Thus, $^{64}$Ge is an $N=Z$ proton-rich unstable nucleus manifesting
 a soft structure with respect to quadrupole and octupole deformations,
 from experimental and theoretical evidences. 
In fact, inclusion of the $\gamma$ deformation improves the $E2$ 
transitions of negative-parity states in $^{68}$Ge \cite{Faessler}. 

The spherical shell model approach could be more appropriate for describing
 various aspects of nuclear structure. 
It is desirable to add the $g_{9/2}$ orbital to the full $pf$ shell
($f_{7/2}, p_{3/2}, f_{5/2}, p_{1/2}$) for studying both positive and negative
 parity states of $^{64}$Ge, but the shell model calculation in this space is 
 impractical at present because of the huge dimension.
So we restrict the model space to the $p_{3/2}, f_{5/2}, p_{1/2}$, and $g_{9/2}$
 orbitals, and carry out the shell model calculation with the recently developed 
 shell model code \cite{Mizusaki}. 
There are few effective shell-model interactions \cite{Caurier} in this model
 space. 

Recently, an extended $P+QQ$ force \cite{Hasegawa} was applied to
 the $f_{7/2}$-shell nuclei. This interaction is schematic but works remarkably well. 
The conventional $P+QQ$ force was first suggested by Bohr and Mottelson,
 and widely used by Kisslinger and Sorensen, Baranger and Kumar, and many authors.
  Unlike the original application to heavy nuclei,
 the extended $P+QQ$ interaction is isospin-invariant \cite{Kaneko}. 
In this Rapid Communication, we introduce the octupole-octupole ($OO$) force 
into the extended $P+QQ$ force model to describe negative-parity states. 
This interaction is quite useful for studying not only the $\gamma$ softness 
but also the octupole instability mentioned above. 
The $QQ$ and $OO$ forces are the long-range and the deformation-driving part of the 
effective interaction. Contrary to this, the monopole pairing force 
can be associated with short-range force, and restores the spherical shape. 
Thus, the competitions among the $QQ$, $OO$, and monopole pairing forces 
are expected to be important for shape transitions of quadrupole and octupole
deformations in $^{64}$Ge. 
The $P+QQ$ force with the $OO$ interaction will be suitable for studying 
the monopole pairing, quadrupole, and octupole correlations. 
Recently, the $fp$ shell model calculation \cite{Honma} with the FPD6
 interaction \cite{FPD6} has been performed 
in $^{64}$Ge as a test case for quantum Monte Carlo diagonalization (QMCD) method. 
The projected shell model calculation \cite{Sun} has been performed in $^{64}$Ge modifying the standard Nilsson parameters. 

Since protons and neutrons in the $N=Z$ nuclei occupy the same levels, one would expect 
strong proton-neutron ({\it p-n}) interactions \cite{Goodman}. 
In particular, one of the most interesting questions in the study of nuclear structure is
what roles the {\it p-n} interaction play in the nuclear deformation. 
The long-range {\it p-n} isoscalar $(T=0)$ 
interaction between valence nucleons has been suggested to be a source of the nuclear 
deformation \cite{Federman}. On the other hand, the isoscalar $QQ$ interaction used in 
the $P+QQ$ force 
model has very strong {\it p-n} component $Q_{p}Q_{n}$, which gives rise to nuclear 
quadrupole deformation \cite{Dobacz}. 
The $Q_{p}Q_{n}$ interaction is expected to be important 
for quadrupole collectivity in $^{64}$Ge. 
The rotational behavior  of $T=0$ and $T=1$ bands in the odd-odd $N=Z$ nucleus 
$^{62}$Ga is recently studied using the spherical shell model and the cranked 
Nilsson-Strutinsky model \cite{Aberg}. 

In order to study the octupole correlation, 
let us introduce an isoscalar octupole interaction $H_{OO}$ with the force strength 
$\chi_{3}$ to the extended $P+QQ$ model\cite{Hasegawa} with monopole 
corrections $H_{\rm m}^{\rm corr}$:
\begin{eqnarray}
H & = & H_{0} + H_{P_{0}} + H_{P_{2}} + H_{QQ} + H_{OO} + H_{\rm m}^{\rm corr} \nonumber \\
 & = & \sum_{\alpha}\varepsilon_{a}c_{\alpha}^{\dag}c_{\alpha}-\sum_{J=0,2}g_{J}\sum_{M\kappa}{P}^{\dag}_{JM1\kappa}{P}_{JM1\kappa} \nonumber \\
  & {} &  -\frac{1}{2}\sum_{M}\chi_{2}:{Q}^{\dag}_{2M}{Q}_{2M}: 
          -\frac{1}{2}\sum_{M}\chi_{3}:{O}^{\dag}_{3M}{O}_{3M}: + H_{\rm m}^{\rm corr}, \nonumber \\
\end{eqnarray}
where 
$\varepsilon_{a}$ is a single-particle energy, 
${P}_{JMT\kappa}$ is the pair operator with angular momentum $J$ and isospin $T$,
and ${Q}_{2M}$ (${O}_{3M}$) is the isoscalar quadrupole (octupole) operator. 
Due to the isospin-invariance, each term of the above Hamiltonian includes {\it p-n}
components, which play important roles in $N=Z$ nuclei. 

We carried out shell model calculations in a model space restricted to 
the $2p_{3/2}, 1f_{5/2}, 2p_{1/2}$, and $1g_{9/2}$ orbitals
 (called $pfg$-shell henceforth).
The model assumes a closed $^{56}_{28}$Ni$_{28}$ core and does not allow
 for core breaking. 
The neutron single-particle energies of $2p_{3/2}, 1f_{5/2}$, $2p_{1/2}$,
 and  $1g_{9/2}$ in this $pfg$-shell region can be read from the low-lying states
 of $^{57}$Ni, because the low-lying states of $^{57}$Ni are well
 characterized as pure single-particle levels when $^{56}$Ni is a closed shell core.
 The adopted single-particle energies relative to the $2p_{3/2}$ are 
$\varepsilon_{p3/2}=0.0$, $\varepsilon_{f5/2}=0.77$, $\varepsilon_{p1/2}=1.11$, and 
$\varepsilon_{g9/2}=3.70$ in MeV \cite{Rudolph2}. 
Since the above Hamiltonian is assumed to be an 
isospin-invariant, the proton single-particle energies are taken as the same values as 
the neutron single-particle energies. 
The force strengths of the extended $P+QQ$ interaction are taken
so as to reproduce the energy levels of low-lying states in $^{64}$Ge as follows: 
\begin{eqnarray}
g_{0}=0.426(42/A), \hspace{0.5cm} g_{2}=0.274(42/A)^{5/3}, \nonumber  \\
\chi_{2}=\chi_{2}^{0}(42/A)^{5/3}/b^{2}=0.567(42/A)^{5/3}/b^{2}, \\
\chi_{3}=\chi_{3}^{0}(42/A)^{2}/b^{6}=0.275(42/A)^{2}/b^{6}, \nonumber
\end{eqnarray}
where $g_{0}$, $g_{2}$, $\chi_{2}$, and $\chi_{3}$ are the monopole pairing, 
quadrupole-pairing, $QQ$, and $OO$ force strengths, respectively.
We adopt the harmonic-oscillator range parameter $b\sim A^{-1/3}$,
the effective charge $e_{p}=1.50e$ for proton and $e_{n}=0.50e$ for neutron. 
We adjust phenomenologically force strengths of several monopole corrections 
so as to approximately reproduce the low-lying energy levels of $^{64}$Ge. 
These force strengths can also reproduce the low-lying energy levels of 
$^{58-66}$Ni, $^{60-64}$Zn, $^{66}$Ge, and $^{68}$Se. 

In Fig. 1, calculated energy spectra are compared with experimental data
 for $^{64}$Ge. 
Two side bands are shown in addition to the ground-state band, i.e., 
 positive-parity band {on the band head $2^{+}$ and negative-parity band
 on the band head $3^{-}$. 
The calculations reproduce the observed three bands at good energies.
 The agreement between theory and experiment for the ground-state
 band up to spin $I=8$ is good. 
The calculated $B(E2)$ value between the ground state and 
the first excited $I=2^{+}$ state is $B(E2;2_{1}^{+}\rightarrow 0_{1}^{+})$=245.3 
$(e^{2}$fm$^{4})$ corresponding to the quadrupole deformation $\beta\sim$0.2. 
This value is consistent
 with the predictions of $\beta\sim$0.22 by M${\rm \ddot{o}}$ller 
and Nix \cite{Moller} and of $\beta\sim$0.22 by Ennis {\it et al.} \cite{Ennis},
 and is comparable to the experimental data 12 W.u. of $^{66}$Ge nucleus. 

\begin{figure}
  \begin{center}
    \leavevmode
    \epsfxsize=8cm
    \epsfysize=7cm

    \epsffile{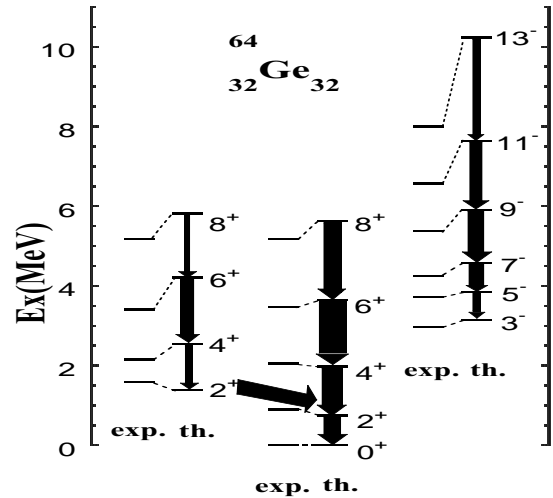}
  \end{center}
  \caption{Comparison of experimental and calculated energy levels of $^{64}$Ge.
           The arrows designate $E2$ transitions with the calculated $B(E2)$
            values indicated by their widths.}
  \label{fig:T=1(N=Z)}
\end{figure}

The calculated occupation numbers of the $p_{3/2},$ $f_{5/2},$ $p_{1/2}$, and $g_{9/2}$
 orbitals in the ground-state band are 
 3.6, 3.2, 0.6, and 0.6 on the average, respectively.
  More than four nucleons are excited from the unperturbed configuration 
  $(p_{3/2})^{8}$. 
The full $fp$ shell model calculation \cite{Honma} with the FPD6 interaction \cite{FPD6}
using the QMCD method has recently been performed for low-lying $I=0_{1}^{+}, 2_{1}^{+},
 2_{2}^{+}$, and $4_{1}^{+}$ states of positive parity in $^{64}$Ge. 
The FPD6 calculation predicts the deformation $\beta_{2}\sim 0.28$ which is somewhat 
 larger than the other predictions $\beta_{2}\sim 0.22$, and gives the triaxiality
 $\gamma\sim 27^{\circ}$ which is consistent with the others predictions. 
 The FPD6 interaction seems too strong to yield appropriate collectivity
  \cite{Honma3}, due to its drawback \cite{Honma2}. 
 This can be seen from 
 the occupation numbers of $f_{7/2}, p_{3/2}, f_{5/2}$, and $p_{1/2}$
 which are 15.1, 2.6, 5.5, and 0.8, respectively.
 Two more nucleons are jumping to the orbitals above $p_{3/2}$ as compared with
 our result. 
 The stronger collectivity caused by the FPD6 interaction is probably
 attributed to the mixture of the} three orbitals ($f_{7/2}$, $p_{3/2}$, and $f_{5/2}$)
 due to the large matrix elements between $(f_{7/2}, p_{3/2})$ and $f_{5/2}$.
 This is the reason why $B(E2;2_{1}^{+}\rightarrow 0_{1}^{+})$ =
$5\times 10^{2}$($e^{2}$fm$^{4}$) obtained in Ref. \cite{Honma} is almost twice 
that of ours 245.3 $(e^{2}$fm$^{4})$.

\begin{figure}
  \begin{center}
    \leavevmode
    \epsfxsize=9cm
    \epsfysize=10cm
    \epsffile{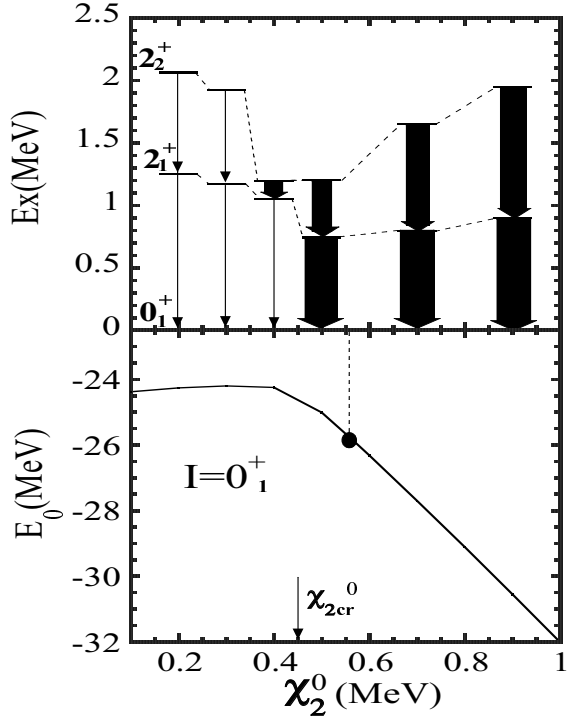}
  \end{center}
  \caption{The excitation energies of the first and second excited $2^{+}$ states
   in the upper figure and the $0^{+}$ ground-state energy in the lower figure,
   as a function of the quadrupole force strength. 
   The arrows indicated by their widths designate $E2$ transitions
   with the calculated $B(E2)$ values.}
\end{figure}

The ratio of excitation energies for the $4_{1}^{+}$ state and the $2_{1}^{+}$ state
gives additional information with respect to the shape.
This ratio is 3.33 for a rigid rotor, 2.0 for a pure vibrator,
 and $\sim$2.4 for a $\gamma$-soft nucleus. Therefore, the ratio 2.65 in the calculation 
 indicates $\gamma$ softness. 
We can also see the $\gamma$-soft nature in $E2$ transitions. 
The $B(E2)$ value of $2_{2}^{+}\rightarrow 2_{1}^{+}$ transition is larger than 
$B(E2;2_{2}^{+}\rightarrow 0_{1}^{+})$, and the ratio $B(E2;2^{\dag}_{2}\rightarrow 
2^{\dag}_{1})/B(E2;2^{\dag}_{2}\rightarrow 0^{\dag}_{1})$ is $\sim$ 27, corresponding to 
$\gamma\sim 26^{\circ}$ in the Davydov model \cite{Davydov}. 
 The present result is in agreement with the triaxiality 
estimated from the experimental data and the other theoretical models, 
though it is difficult for the shell model calculation to definitely discuss 
$\gamma$ softness and triaxiality. 

On the other hand, the low-lying negative-parity states with $I=3^{-},5^{-},7^{-},$
and $9^{-}$ are nicely reproduced. 
The octupole force newly introduced to our model plays an essential role
 in describing the negative-parity states. 
The value of $B(E2;5^{-}\rightarrow 3^{-})$ is small, 
while the $B(E2)$ values between higher states above $3^{-}$ are large.
This indicates a different structure of $3^{-}$ from the higher
 negative-parity states. 
 The same is known from the occupation numbers of $p_{3/2}, f_{5/2}, p_{1/2},$
 and $g_{9/2}$ orbitals. Their values of the $I=3^{-}$ state 
are, respectively, 4.0, 1.2, 0.6, and 2.2, while the average values of
the negative-parity states above $3^{-}$ are 3.4, 2.7, 0.5, and 1.4. 
More nucleons are jumping from the unperturbed configuration
 $(p_{3/2})^{7}(g_{9/2})^{1}$ in the $3^{-}$ state as compared with the states above
 $3^{-}$. 
The $I=3^{-}$ state is mainly a coherent mixture of neutron and proton
 $(p_{3/2}f_{5/2}g_{9/2})_{3^{-}}$ configurations. 
The present result is consistent with the calculations by Petrovici
 and Faessler \cite{Faessler}. 
In Fig. 1, the calculated $11^{-}$ and $13^{-}$ levels are
 considerably higher than the observed excitation energies. 
This may suggest that excitation from the $f_{7/2}$ orbital in $^{56}$Ni core
should be taken into account for these states. 

Figure 2 shows the excitation energies of the first and second $2^{+}$ states
 and the $0^{+}$ ground-state energy as a function of the quadrupole force strength 
$\chi_{2}^{0}$. 
The other force strengths are fixed to the values of Eq. (2). 
 We can see a level crossing of the first excited $2_{1}^{+}$ state and 
the second excited $2_{2}^{+}$ state near the crossing strength
 $\chi_{2{\rm cr}}^{0}\sim$0.45 MeV. 
When the quadrupole force strength is apart from $\chi_{2{\rm cr}}^{0}$, 
the second excited $I=2_{2}^{+}$ state rises abruptly in energy,
while the first excited $I=2_{1}^{+}$ state does not change largely. 
This means 
that $\gamma$ softness or triaxiality is realized beyond the crossing point. 
The ground-state energy rapidly decreases beyond this crossing point, while 
 it is almost constant for $0 < \chi_{2}^{0} < 0.45$ MeV. 

\begin{figure}
  \begin{center}
    \leavevmode
    \epsfxsize=8cm
    \epsfysize=7cm
    \epsffile{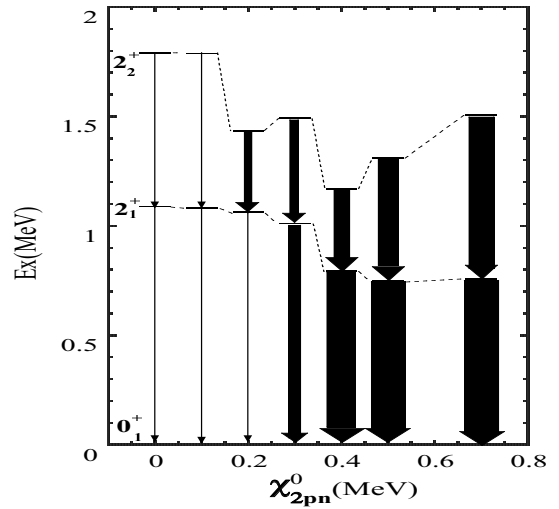}
  \end{center}
  \caption{Excitation energies of the first and second excited $2^{+}$ states 
  as a function of the {\it p-n} quadrupole force strength $\chi_{2pn}^{0}$. 
   The arrows designate $E2$ transitions with the calculated $B(E2)$
   values indicated by their widths.}
\end{figure}

There seems to be a critical point of phase 
transition with respect to the quadrupole correlation. 
Indeed the $B(E2;2_{1}^{+}\rightarrow 0_{1}^{+})$ values dramatically increase when the 
quadrupole force strength goes beyond the critical point. 
The force strength adopted in Fig. 1 is $\chi_{2}^{0}$=0.567 MeV denoted by a 
solid circle in the lower part of Fig. 2, and its position lies in the deformed 
region but near the critical point. 
We can also see that the $B(E2;2_{2}^{+}\rightarrow 2_{1}^{+})$ value
is large for $\chi_{2}^{0} > 0.45$ MeV. 
This is consistent with the $\gamma$ softness or the 
triaxiality discussed above. 

\begin{figure}
  \begin{center}
    \leavevmode
    \epsfxsize=9cm
    \epsfysize=10cm
    \epsffile{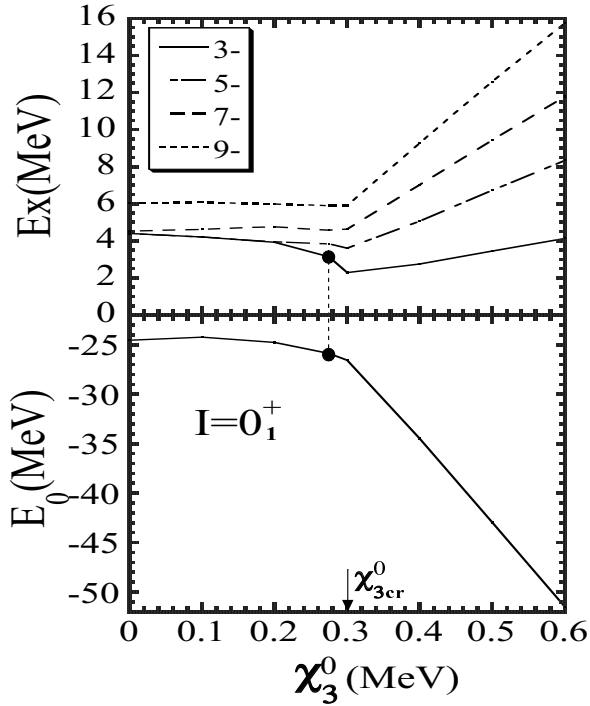}
  \end{center}
  \caption{Excitation energies of the negative-parity states
  (in the upper figure) and the $0^{+}$ ground-state energy 
  (in the lower figure)
  as a function of the octupole force strength.
   The solid circles show the energies for the adopted force strength.}
  \label{fig:T=1(N=Z)}
\end{figure}

   As mentioned earlier, it is very interesting to study
 roles of the proton-neutron part of the $QQ$ force ($Q_{p}Q_{n}$).
 We make the study by varying the strength of $Q_{p}Q_{n}$
 ($\chi_{2pn}^{0}$) and keeping the other force strengths of Eq. (2).
 This is effective in seeing the dependence of quadrupole deformation
 on $Q_{p}Q_{n}$, though the Hamiltonian without the isovector type of $QQ$ force stops
 beeing isospin invariant.
 Figure 3 shows the excitation energies of the first and second
 $2^{+}$ states and the $B(E2)$ values as a function of $\chi_{2pn}^{0}$.
The first $I=2_{1}^{+}$ state is flat in energy with respect to the force strength 
$\chi_{2pn}^{0}$. However, the second excited $I=2_{2}^{+}$ state strongly depends on 
$\chi_{2pn}^{0}$, and becomes lowest near $\chi_{2pn}^{0}\sim 0.35$ MeV. 
In the isoscalar $QQ$ force, $\chi_{2pn}^{0}$ is equal to the magnitude of proton-proton 
({\it pp}) and neutron-neutron ({\it nn}) force strengths due to 
the isospin-invariance, i.e., $\chi_{2pn}^{0}=\chi_{2pp}^{0}=\chi_{2nn}^{0}$. 
The force strength $\chi_{2pn}^{0}=0.567$ MeV which is adopted in Fig. 1
 leads to the large $B(E2)$ value 
$B(E2;2_{1}^{+}\rightarrow 0_{1}^{+})$=245.3 $(e^{2}$fm$^{4})$ 
corresponding to the quadrupole deformation $\beta\sim$0.2, as mentioned above. 
We can see, however, that $B(E2;2_{1}^{+}\rightarrow 0_{1}^{+})$ is very small 
 for $0 < \chi_{2pn}^{0} < 0.35$ MeV. This suggests that a phase transition occurs
 near the critical point $\chi_{2pn{\rm cr}}^{0}\sim 0.35$ MeV. 
The $B(E2)$ value of the $2_{2}^{+}\rightarrow 2_{1}^{+}$ transition also
 becomes large for $\chi_{2pn}^{0} > 0.35$ MeV, which results in the expected
 triaxiality $\gamma\sim 26^{\circ}$ when $\chi_{2pn}^{0}=0.567$ MeV. 
Thus, the $Q_{p}Q_{n}$ interaction plays an important role for the $\gamma$ softness or 
the triaxiality in $^{64}$Ge. 

Let us lastly study the negative-parity states as a 
function of the octupole force strength in Fig. 4. 
The other force strengths are fixed to those of Eq. (2). 
As mentioned above, 
the $I=3_{1}^{-}$ state is very collective, 
and the excitation energy of the $I=3_{1}^{-}$ state
 decreases as the octupole force strength increases until $\chi_{3}^{0}\sim$0.3 MeV,
 and increases as it goes beyond this point. 
 The other negative-parity states have an insignificant dependence
  with respect to the force strength $\chi_{3}^{0}$ until the critical point,
  and increase for $\chi_{3}^{0}>$ 0.3. 
 The ground-state energy is almost constant
  for $0 < \chi_{\rm 3cr}^{0} < 0.3$ MeV and decreases quickly when going beyond this 
critical point. It seems that a phase transition occurs near the critical force
 strength 
$\chi_{\rm 3cr}^{0}\sim $0.3 MeV. The octupole force strength 
$\chi_{3}^{0}$=0.275 MeV adopted in Fig. 1 is very close to the critical point 
$\chi_{3{\rm cr}}^{0}$. 
Thus, $^{64}$Ge seems to be near an octupole instability. 

In summary, we have studied quadrupole and octupole correlations
in the even-even $N=Z$ nucleus $^{64}$Ge by means of spherical shell 
model calculations. The $P+QQ$ force model including octupole interaction
and monopole corrections, which is schematic but realistic, 
 is adopted for describing the quadrupole and octupole correlations. 
It is shown that $^{64}$Ge is an unstable nucleus with respect to both the quadrupole 
and octupole deformations. The present results reveal that 
the {\it p-n} $QQ$ interaction ($Q_{p}Q_{n}$) induces an onset of quadrupole 
deformation and $\gamma$ softness. It can be expected to 
play an important role in the prolate-oblate shape coexistence of the
 neighboring even-even $N=Z$ nucleus $^{68}$Se, which has been recently
  observed \cite{Skoda,Fischer}.

\bibliographystyle{unsrt}

\end{document}